\documentclass[twocolumn,prb]{revtex4}
\usepackage{graphicx} 
\usepackage{subfigure} 
\usepackage{bm} 
\usepackage{amssymb} 
\usepackage{wasysym} 
 
\newcommand{\Ket}[1]{\vert  #1 \rangle} 
 
\newcommand{\MatEl}[3]{\langle  #1 \vert #2 \vert #3\rangle} 
\newcommand{\Amp}[2]{\langle  #1 \vert  #2 \rangle} 
 
\newcommand{\Avg}[1]{\langle  #1  \rangle}

\renewcommand{\phi}{\varphi} 
\renewcommand{\epsilon}{\varepsilon} 
\renewcommand{\vec}[1]{{\bf #1}} 
 
\begin{document} 
 
\title{Phase Transitions in Dissipative Quantum Transport and Mesoscopic\\ Nuclear Spin Pumping} 
\author{M. S. Rudner$^{1}$ and L. S. Levitov$^{2}$} 
\affiliation{ 
$^{(1)}$ Department of Physics, 
 Harvard University, 17 Oxford St., 
 Cambridge, MA 02138\\ 
$^{(2)}$ Department of Physics, 
 Massachusetts Institute of Technology, 77 Massachusetts Ave, 
 Cambridge, MA 02139} 
 
\begin{abstract} 
Topological phase transitions can occur in the dissipative dynamics of a quantum system when the ratio of matrix elements for competing transport channels is varied.  
Here we establish a relation between such behavior 
in a class of non-Hermitian quantum walk problems [M. S. Rudner and L. S. Levitov, Phys. Rev. Lett. {\bf 102}, 065703 (2009)] and nuclear spin pumping in double quantum dots, which is mediated by the decay of a spin-blockaded electron triplet state 
in the presence of spin-orbit and hyperfine interactions. 
The transition occurs when the strength of spin-orbit coupling exceeds the strength of the net hyperfine coupling, and results in the complete suppression of nuclear spin pumping. 
Below the transition point, nuclear pumping is accompanied by a strong reduction in current
due to the presence of non-decaying ``dark states'' in this regime.
Due to its topological character, the transition is expected to be robust against dephasing of the electronic degrees of freedom. 
\end{abstract} 
 
\maketitle 
 
Since the first observation of spin blockade in vertical GaAs double quantum dots~\cite{OnoSB}, spin-blockaded transport has been observed in a variety of systems such as lateral double quantum dots 
in GaAs, Si, and Si/SiGe heterostructures~\cite{Koppens,Liu08,Shaji08},
InAs nanowires~\cite{Pfund,Frolov}, and carbon nanotubes~\cite{Churchill}. 
Much of this work was driven by the need to better understand the coupled dynamics of electron and nuclear spins in double dot systems with potential applications in spintronics and quantum computation. 
A variety of interesting and surprising phenomena such as current bistabilities and hysteresis~\cite{OnoTarucha,Koppens,Pfund,Churchill}, very long time scale switching~\cite{Koppens}, and periodic oscillations~\cite{OnoTarucha} have been observed, and their origins linked to the dynamical polarization of nuclear spins (DNP). 
Although the involvement of nuclear spins in these phenomena is clear, in many cases the underlying mechanisms remain a mystery. 
 
\begin{figure} 
\includegraphics[width=3.2in]{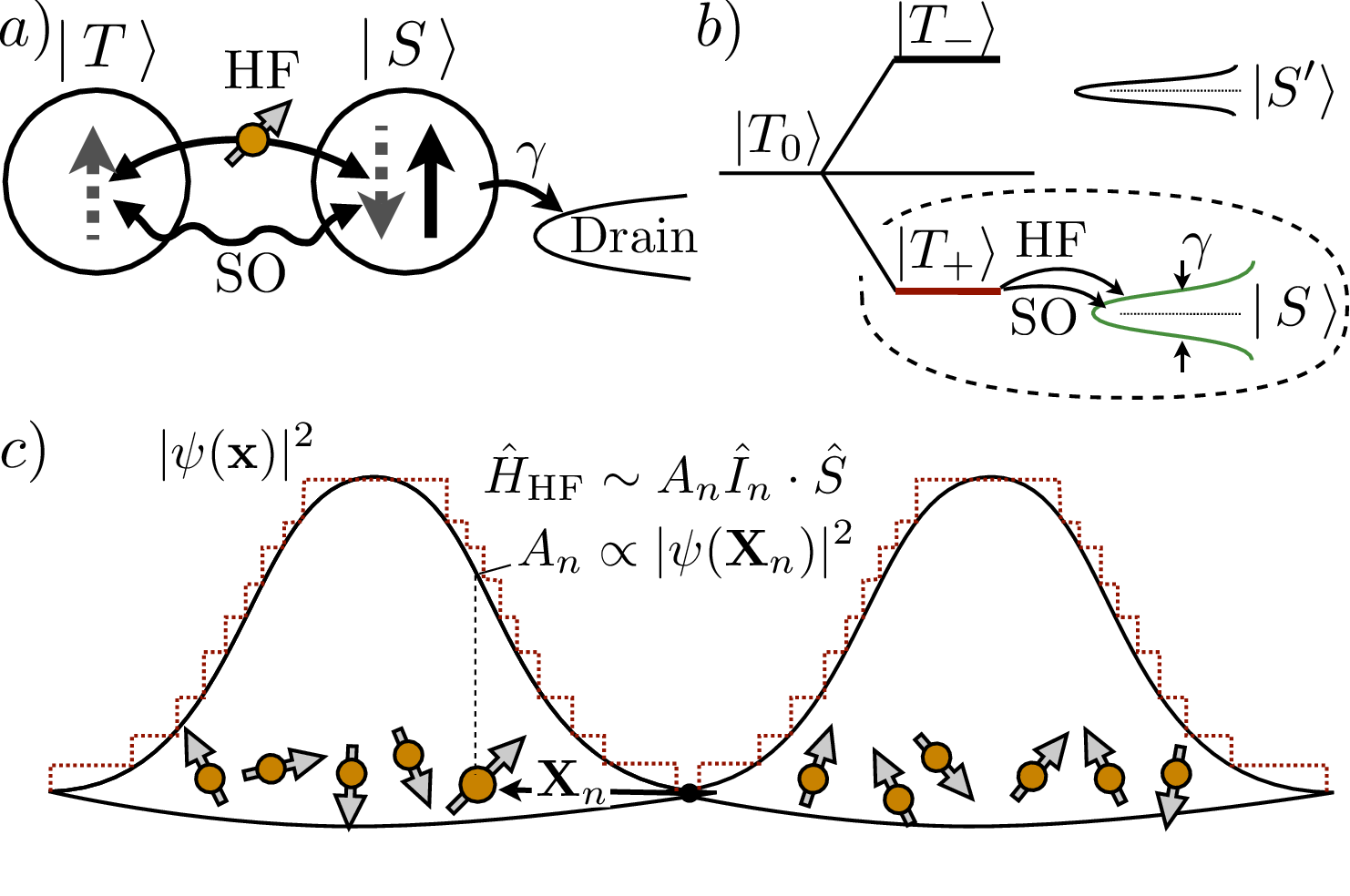} 
 \caption[]{Competition between hyperfine and spin-orbit decay in spin-blockaded double quantum dots. 
a)  
A triplet state decays 
via a hyperfine or spin-orbit mediated transition to a singlet state. 
The singlet state is coupled to the drain, and decays with rate $\gamma$. 
b) 
Energy levels and transitions. 
We focus on the subspace 
indicated by the dashed oval.
c) Inhomogeneous hyperfine coupling due to non-uniform electron density. 
Red dotted line shows approximation to smooth density profile consisting of uniform density shells in which nuclear spins collectively couple to an electron as
separately conserved ``giant spins.'' 
 } 
\label{Cartoons} 
\vspace{-4mm} 
\end{figure}

Spin blockade of dc transport occurs in a two-electron double quantum dot when the electron spins form a triplet state that prohibits both electrons from occupying the same site. 
In this case, as shown in Fig.\ref{Cartoons}a, residual current arises from mechanisms that do not conserve the electron spin such as 
the spin-orbit interaction, and the hyperfine coupling to nuclear spins in the host lattice. 
Because the hyperfine contact interaction conserves the total spin of all electrons plus nuclei, each hyperfine-mediated electron spin flip is accompanied by a nuclear spin flip in the opposite direction. 
In the presence of strong spin-orbit coupling, e.g. as in InAs systems~\cite{Pfund,Frolov}, the physics of DNP  
can be very different from that studied previously in the absence of spin-orbit coupling\cite{RudnerDNP, RudnerCooling}. 
In particular, by making several transitions 
between singlet and triplet states using a combination of hyperfine and spin-orbit processes, the decay of a single electron spin can lead to a change of nuclear polarization by an amount which can have either sign, and a magnitude potentially even greater than one unit of angular momentum\cite{Larmor}.  
 
 
In this paper we explore the rich quantum dynamics of coupled electron and nuclear spins that results from the coherent competition of hyperfine and spin-orbit decay channels in spin-blockaded quantum dots. 
We focus on the polarization transferred to the nuclear spin bath by the decay of a single electron in one of the blockaded triplet states. 
This quantity reveals a strikingly high sensitivity of the DNP production efficiency to the presence of the competing spin-orbital decay channel. 

In order to study DNP in this regime,
we develop a class of models which capture the essential physics of polarization transfer during electron spin decay. 
One of the main difficulties in describing nuclear pumping in the presence of spin-orbit coupling stems from the lack of a concrete conservation law which directly relates the changes of electron and nuclear spin polarizations. 
Any model of this process must account for the possibility of multiple electron spin transitions which can lead to a variety of final nuclear spin states. 
For a typical system containing $N \approx 10^6$ nuclear spins, the exponentially large Hilbert space makes exact analytical or numerical solutions difficult to obtain. 
However, as we show below,
the problem can be made tractable by introducing approximations which greatly reduce the number of variables while retaining the key degrees of freedom responsible for the mechanism of polarization transfer. 
 
\begin{figure} 
\includegraphics[width=3.2in]{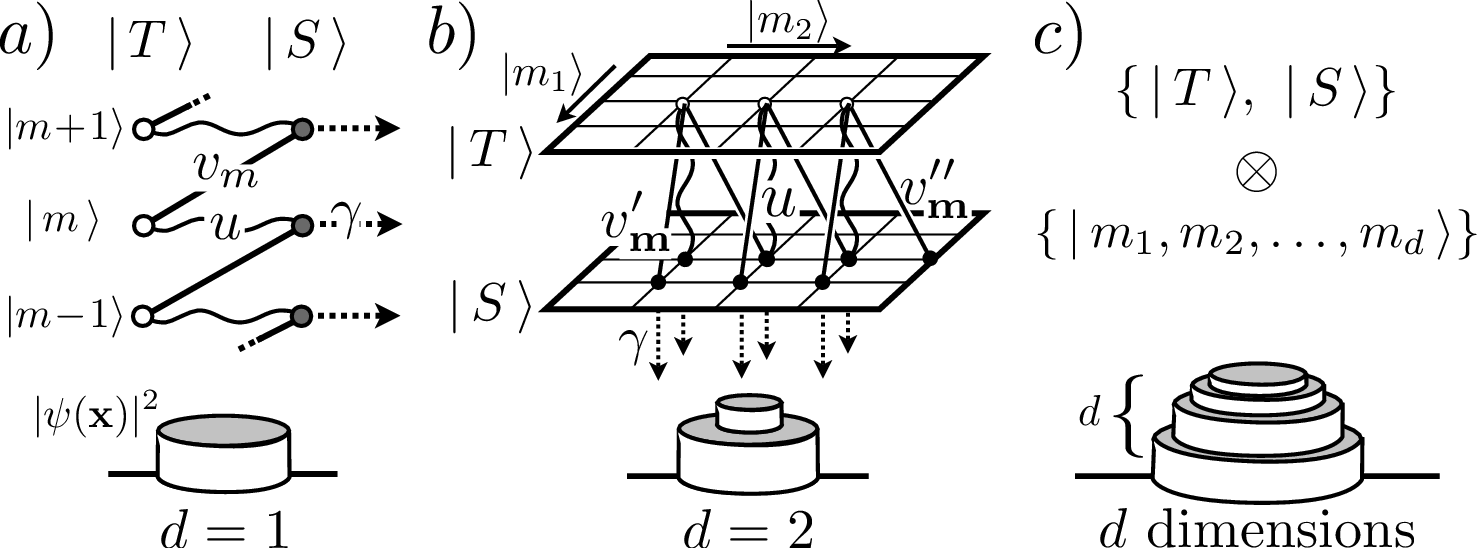} 
 \caption[]{Mapping of coupled dynamics of electron and nuclear spins onto quantum walk models describing 
single particle hopping on suitable lattices.
a) In the uniform coupling model, the state of the conserved ``giant-spin'' is labeled by its $z$-component $m$. 
Spin-orbit and hyperfine transitions with amplitudes $u$ and $v_m$ can be interpreted as hopping in a tight-binding model defined
on a one-dimensional bipartite lattice. 
b) For 
two 
groups of nuclei, 
the nuclear state is labeled by the $z$-components $m_1$ and $m_2$ of two collective spins $I_1$ and $I_2$, 
and the dynamics are described 
by a quantum walk on a two-dimensional bipartite lattice. 
c) For an 
electron density profile approximated by $d$ constant density shells, the dynamics can be viewed in terms of a quantum walk on a $d$-dimensional bipartite lattice 
with hopping amplitudes $v^{(\alpha)}$ describing hyperfine transitions with each of the $\alpha=1\ldots d$ collective spins. 
 } 
\label{MultiDQW} 
\vspace{-4mm} 
\end{figure}

We begin by employing the ``giant-spin'' model, as used e.g. in Ref.[\onlinecite{Taylor03}], in which the electrons interact with a single large collective spin formed from all the spins in the nuclear spin bath. 
This model describes the case where the local hyperfine coupling to each nuclear spin within each dot takes on a uniform value, $\bar{A}$.  
Within this approximation, the problem near the singlet-triplet resonance, circled in Fig.\ref{Cartoons}b, can be viewed as a one-dimensional hopping problem in the space of polarization of the giant collective nuclear spin, see Fig.\ref{MultiDQW}a. 
We study this model numerically, and obtain additional insight from comparison 
to the solution of a related quantum walk model\cite{1DPRL} 
in which the quantity analogous to DNP is described by a topological invariant which takes on integer values.
We find a non-analytic dependence of the polarization transfer on the ratio of hyperfine and spin-orbit coupling strengths, with complete suppression of DNP when the spin-orbit coupling exceeds the net transverse hyperfine field.  This behavior is a direct manifestation of the topological phase transition which occurs in the 
quantum walk model\cite{1DPRL}.

To investigate the role of inhomogeneous hyperfine coupling, we then employ a model in which the nonuniform hyperfine couplings are approximated by $d$ shells of constant coupling, as shown in Fig.\ref{Cartoons}c. 
Here, nuclear spins couple to form $d$ large collective spins which interact with the electrons. 
The resulting dynamics can be viewed as a hopping problem in a $d$-dimensional space indexed by the polarizations of each of the $d$ collective spins, see Figs.\ref{MultiDQW}b and c.  
  
 
Although a numerical approach is not possible for the general case, exact analytic results for a related $d$-dimensional hopping problem show universal features which are independent of the specific grouping of nuclear spins. 
The behavior obtained for $d>1$ is essentially analogous to that found in the $d=1$ case (the giant spin model).
While the details of the behavior near the transition are sensitive to the 
particular decomposition into collective spins, the strong suppression of DNP in the spin-orbit dominated phase is found to be generic, suggesting that it will persist in more realistic models.


 
\section{Construction of the Model} 
 
 
We begin by reviewing the relevant two-electron states of a spin-blockaded double quantum dot, see e.g. Ref.[\onlinecite{OnoSB}]. 
For simplicity, suppose that the left and right dots each support a single orbital state $\Ket{L}$ or $\Ket{R}$. 
An applied potential bias approximately compensates the charging energy when two electrons occupy the right dot. 
Energetically, electrons which form a spin singlet can thus assume either the $(1, 1)$ or $(0, 2)$ charge configuration, with one electron on each dot or both electrons on the right dot: 
\begin{eqnarray} 
\label{BareSinglets} 
 \nonumber \Ket{(1,1)_S} &=& \frac12\left(\Ket{L\,R} + \Ket{R\,L}\right)\otimes\left(\Ket{\uparrow\downarrow} - \Ket{\downarrow\uparrow}\right)\\ 
  \Ket{(0,2)_S} &=& \frac{1}{\sqrt{2}}\Ket{R\,R}\otimes\left(\Ket{\uparrow\downarrow} - \Ket{\downarrow\uparrow}\right). 
\end{eqnarray} 
When the electrons form a spin triplet, the Pauli exclusion principle forbids double occupation of the right dot. 
 
Due to tunnel coupling, the singlet states $\Ket{(1,1)_S}$ and $\Ket{(0,2)_S}$, Eq.(\ref{BareSinglets}), hybridize to form ``bonding'' and ``anti-bonding'' states $\Ket{S}$ and $\Ket{S'}$. 
In the presence of an external magnetic field, the triplet splits into its three Zeeman sublevels. 
For concreteness, we consider decay of the state $\Ket{T_+}$ when its energy is close to that of the singlet states $\Ket{S}$, with all other states far away in energy (see Fig.\ref{Cartoons}b). 
The relevant electronic states for our model are thus 
\begin{eqnarray} 
\label{States} 
\nonumber \Ket{T_+} &=& \frac{1}{\sqrt{2}}\left(\Ket{L\,R} - \Ket{R\,L}\right)\otimes\Ket{\uparrow\uparrow}\\ 
 \Ket{S} &=& C_{11}\Ket{(1,1)_S} + C_{02}\Ket{(0,2)_S}. 
\end{eqnarray} 
 
Because of spin-orbit coupling, interdot tunneling is accompanied by a spin-rotation which couples the triplet and singlet states $\Ket{T_+}$ and $\Ket{S}$ with an amplitude $u$. 
In the basis of Eq.(\ref{States}), the purely electronic part of the Hamiltonian is written as 
\begin{eqnarray} 
\label{H0}\hat{H}_0 = \left(\begin{array}{cc} \epsilon_{T_+} & u\\u & \tilde{\epsilon}_S\end{array}\right), \quad \tilde{\epsilon}_S = \epsilon_S - i\gamma/2, 
\end{eqnarray} 
where $\epsilon_{T_+}$ and $\epsilon_S$ are the energies of the states $\Ket{T_+}$ and $\Ket{S}$, respectively. 
The imaginary term $-i\gamma/2$ in $\tilde{\epsilon}_S$ accounts for the decay of the singlet state due to coupling of $\Ket{(0,2)_S}$ to the drain, see Fig.\ref{Cartoons}a. 
Without loss of generality, we take the spin-orbit coupling matrix element $u$ to be real. 
A microscopic derivation of the value of $u$ is beyond the scope of this work. 
However, because the spin rotation occurs during tunneling, $u$ is proportional to the admixture of $\Ket{(0,2)_S}$ in $\Ket{S}$, i.e. to the parameter $C_{02}$ in Eq.(\ref{States}). 
In addition, the value of $u$ is sensitive to the orientation of the dots relative to the crystallographic axes, and to the direction of the applied magnetic field. 
 
The hyperfine interaction between electron and nuclear spins also couples the triplet and singlet states. 
In a two-electron system, the hyperfine Hamiltonian  
\begin{eqnarray} 
  \label{HFHam}  \hat{H}_{\rm HF} = A\sum_n \vec{I}_n\cdot\left[\vec{S}_1\delta(\vec{x}_1 - \vec{X}_n) + \vec{S}_2\delta(\vec{x}_2 - \vec{X}_n)\right] 
\end{eqnarray} 
couples the spin $\vec{S}_{1(2)}$ of each electron to each nuclear spin $\vec{I}_n$, with weight proportional to the probability to find the electron at the location $\vec{X}_n$ of nucleus $n$. 
Here, for simplicity, we consider a single species of nuclear spin, but a generalization to multiple species is straightforward. 
The spin-flip terms $\hat{S}_{1(2)}^\pm \hat{I}_n^{\mp}$ couple the electron states with spin projections along the $z$-axis differing by one unit of angular momentum. 

In the basis of Eqs.(\ref{States}) and (\ref{H0}), Hamiltonian (\ref{HFHam}) takes the form 
\begin{eqnarray} 
  \label{BGHam2x2} \hat{H}_{\rm HF} = \left(\begin{array}{cc} 
     \frac12 \sum_n A_n \hat{I}_n^z  & \frac{C_{11}}{2\sqrt{2}} \sum_n \eta_n A_n \hat{I}_n^-\\ 
   \frac{C_{11}}{2\sqrt{2}} \sum_n \eta_n A_n \hat{I}_n^+ & 0 
    \end{array} 
\right), 
\end{eqnarray} 
where  
\begin{equation} 
  A_n = A\rho(\vec{X}_n) 
\end{equation}  
is the hyperfine coupling weighted by the local electron density $\rho(\vec{X}_n) = \MatEl{\psi}{\delta(\hat{\vec{x}} - \vec{X}_n)}{\psi}$ (see Fig.\ref{Cartoons}c), and, due to antisymmetry of the wavefunction, $\eta_n = +1(-1)$ if nucleus $n$ is located in the left (right) dot. 
Because the hyperfine interaction is local, the off-diagonal matrix elements of Hamiltonian (\ref{HFHam}) between $\Ket{T_+}$ and $\Ket{S}$ are proportional to the amplitude of $\Ket{(1,1)_S}$ in $\Ket{S}$, i.e. to the parameter $C_{11}$ in Eq.(\ref{States}). 
 
The sign factors $\eta_n$ 
indicate that the {\it difference} between transverse nuclear polarizations in the left and right dots couples the electron triplet and singlet levels. 
The mathematical annoyance of alternating signs can be removed by applying a $\pi$-rotation about the $z$-axis to all spins in the right dot via the operator $\hat{U} = e^{-i \pi \sum_{R} \hat{I}^z_n}$, where the sum is taken over all spins in the right dot. 
In the rotated frame, the Hamiltonian $\hat{H}'_{\rm HF} = \hat{U}^\dag \hat{H}_{\rm HF} \hat{U}$ takes the simpler form: 
\begin{eqnarray} 
  \label{nucRotatedSD} \hat{H}'_{\rm HF} = \left(\begin{array}{cc} 
    \frac12 \sum_n A_n \hat{I}_n^z  & \frac{C_{11}}{2\sqrt{2}} \sum_n A_n \hat{I}_n^-\\ 
   \frac{C_{11}}{2\sqrt{2}} \sum_n A_n \hat{I}_n^+ & 0 
    \end{array} 
\right). 
\end{eqnarray} 
The factor $1/\sqrt{2}$ arises from the normalization of the states in Eqs.(\ref{BareSinglets}) and (\ref{States}). 
Up to these numerical prefactors, 
the transformed Hamiltonian is equivalent to that describing the hyperfine interaction for a single electron in a quantum dot with an electron density profile consistent with the distribution of couplings $\{A_n\}$.  
 
\section{Giant Spin Model ($d = 1$)\label{sec:giantSpin}} 
 
In the special case where $A_n = \bar{A}$ for all $n$, which corresponds to an electron density that is uniform within the dots and zero outside, the square of the total nuclear spin operator, $\hat{I}^2 = (\sum_n \vec{I}_n)^2$, commutes with the Hamiltonian. 
In this case, all nuclei in the system act together coherently 
as one ``giant'' spin. 
For fixed $I$, the configuration space of the system is then defined by the electron 
states $\Ket{T_+}$ and $\Ket{S}$, and by the z-projection $m$ of total nuclear spin, $\hat{I}^z\Ket{m} = m\Ket{m}$, with $-I \le m \le I$. 
Combining Eqs.(\ref{H0}) and (\ref{nucRotatedSD}), the Hamiltonian for this system can be written as 
\begin{equation} 
\label{GS}H_{\rm 1D} = \left(\begin{array}{cc}\Delta\epsilon + \frac12 \bar{A} \hat{I}^z & u + \bar{A}\hat{I}^-\\u + \bar{A}\hat{I}^+ & -i\gamma/2\end{array}\right), 
\end{equation} 
where $\Delta\varepsilon$ is the triplet-singlet detuning, and $\hat{I}^{+(-)}$ is the raising (lowering) operator for the giant spin. 
 
Below we neglect the polarization-dependent Overhauser shift $\bar{A} \hat{I}^z$ by absorbing its mean value into the definition of $\Delta\varepsilon$. 
For polarizations which are not too large, this approximation is justified by the fact that the typical off-diagonal matrix elements of $H_{\rm 1D}$ are of the order $\bar{A}\sqrt{N}$, 
while the Overhauser shift only changes by an amount of order $\bar{A}$ when the nuclear polarization changes by one unit of angular momentum. 
For a typical dot containing $N \approx 10^6$ nuclear spins, the variation of the Overhauser shift thus imposes only a small perturbation on the dynamics of the system. 
In a similar spirit, we also ignore the nuclear Zeeman energy, which is assumed to be small compared with the inverse lifetime of the blockaded state.

Decay of the blocked triplet state occurs through electron spin-flip transitions to the state $\Ket{S}$, which is broadened due to its coupling to the drain lead. 
These transitions can be mediated by either the hyperfine interaction or the spin-orbital interaction. 
The hyperfine process is accompanied by a change of the $z$-projection of nuclear spin, $\Delta m=\pm 1$, 
whereas for the spin-orbital process $\Delta m=0$.  
As illustrated in Fig.\ref{MultiDQW}a, the resulting coherent dynamics in the combined Hilbert space of electron and nuclear degrees of freedom can thus be viewed as a hopping problem on a one-dimensional bipartite lattice. 
 
\begin{figure} 
\includegraphics[width=3.25in]{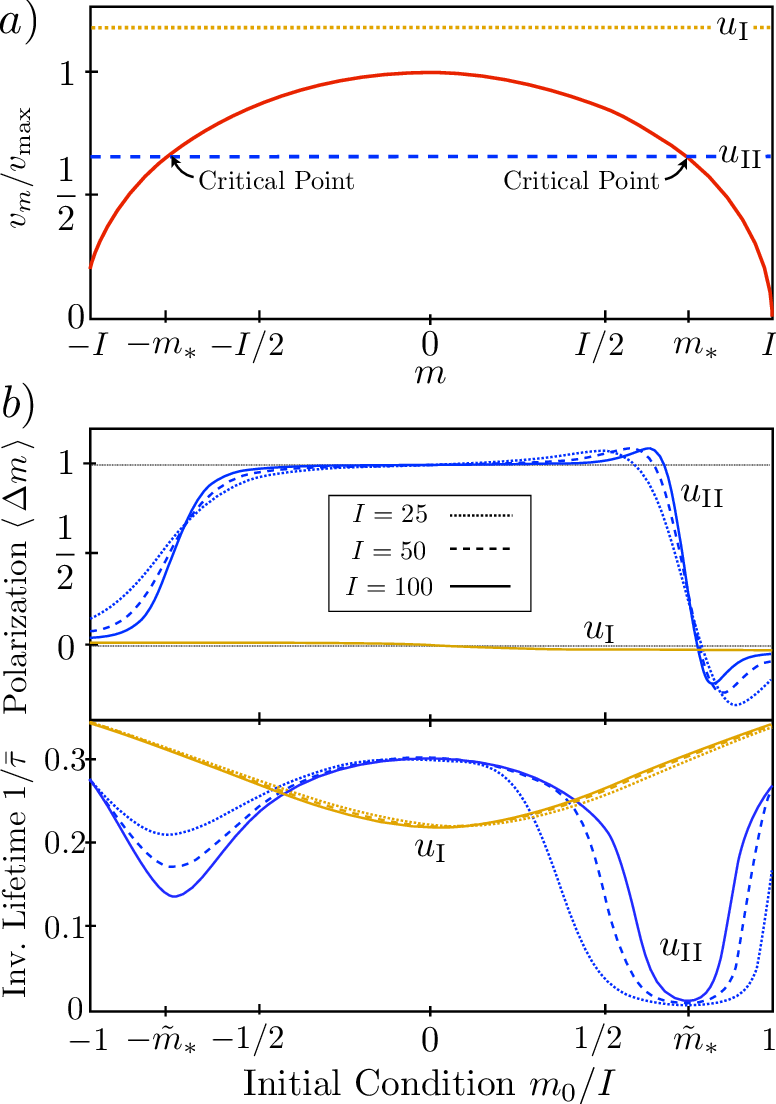}
 \caption[]{ 
Results for the giant spin model, Eq.(\ref{EOM1d}). 
a) Dependence of hyperfine matrix element $v_m$ on the $z$-projection of total nuclear spin for a giant spin of length $I$. 
For moderate values of the spin-orbit coupling, $u = u_{\rm II} < v_{\rm max}$, the system possesses critical points $\pm m_*$, see Eq.(\ref{m*}), where the spin-orbit and hyperfine couplings are roughly equal. 
b) Average polarization change and inverse lifetime versus initial polarization. 
For all traces we take $\Delta\varepsilon = 0$ and $\gamma = v_{\rm max}$. 
When spin-orbit coupling is strong, $u = u_{\rm I} = 1.2 v_{\rm max}$, nuclear spin pumping is suppressed. 
For moderate values of the spin-orbit coupling, $u = u_{\rm II} = 2/3v_{\rm max}$,  $\Avg{\Delta m} \approx 1$ for initial states with $v_{m_0} > u_{\rm II}$, and $\Avg{\Delta m} \approx 0$ for $v_{m_0} < u_{\rm II}$. 
Near the critical 
 points $\tilde m=m_*/I$, the system possesses ``dark states'' with extremely long lifetimes. 
We include a decay term $-i\gamma_T/2$ with $\gamma_T = 10^{-3}\gamma$ in the triplet energy  to cut off the divergence of the lifetime near the critical points. 
} 
\label{Simulation} 
\vspace{-4mm} 
\end{figure}

In the basis $\{\Ket{m}\otimes\Ket{T_+/S}\}$, the state of the system $\Ket{\psi}$ is described by the amplitudes $\psi_m^T = \Amp{m\, T_+}{\psi}$ and $\psi_m^S = \Amp{m\, S}{\psi}$, and evolves according to the equations of motion (with $\hbar = 1$) 
\begin{eqnarray} 
  \label{EOM1d}  
\begin{array}{rcrcccl} 
i  \dot{\psi}^T_{m} &=& \Delta\varepsilon \psi^T_{m} & + & {u} \psi^S_{m} & +&  v_m \psi^S_{m+1}\\ 
i  \dot{\psi}^S_{m} &=& -i(\gamma/2) \psi^S_{m} &+& {u} \psi^T_{m}  &+& v_{m-1} \psi^T_{m-1}, 
\end{array} 
\end{eqnarray} 
with  
\begin{equation} 
\label{vm}v_m = \bar{A}\sqrt{I(I+1) - m(m+1)}. 
\end{equation} 
The hopping amplitudes $v_m$, which originate from the transverse hyperfine field, attain a maximum value $v_{\rm max} = \bar{A}\sqrt{I(I+1)}$ for unpolarized states $m \approx 0$, and become small near maximum polarization $|m| \lesssim I$, see Fig.\ref{Simulation}a. 
 
Suppose the system is initially in the blockaded electron spin state, with nuclear polarization $m_0$. 
What is the average change in nuclear polarization $\Delta m = m - m_0$ caused by the decay of the electron spin?  
To formulate the problem more precisely, we consider the situation where an electron is injected into the triplet state $\Ket{T_+}$ at time $t = 0$, with an initial nuclear spin state characterized by total angular momentum $I$ and $z$-projection $m_0$. 
The system then executes a ``quantum walk'' under the equations of motion  (\ref{EOM1d}), with initial state 
\begin{eqnarray} 
  \label{IC1d} \psi^T_{m} = \delta_{m,m_0}, \quad  \psi^S_{m} = 0.  
\end{eqnarray} 
The wave packet describing the quantum walker will spread throughout the lattice and leak out through its components on the $S$-sites, 
decaying completely as $t\rightarrow\infty$.  
The value of $m$ at the site from which the system decays determines the final value of nuclear polarization left behind when the electron escapes. 
Given the probability $P_{m}$ for the system to decay from each singlet site $m$, we would like to evaluate 
\begin{eqnarray} 
  \label{DeltaM1d}  \Avg{\Delta m} \equiv \sum_m (m - m_0)\, P_{m}  
,\ P_{m}=\int_0^\infty \gamma | \psi^S_{m}(t)|^2 \, dt. 
\end{eqnarray} 
This expression for $P_{m}$ results from the fact that the non-Hermitian equations of motion (\ref{EOMd}) lead to decay which is a sum over {\it local} terms describing decay from each site of the lattice, $\frac{d}{dt}\Amp{\psi}{\psi} = -\sum_m \gamma \vert \psi^S_m\vert^2$. 
Because the system decays completely as $t\rightarrow\infty$, $\sum_m P_m = 1$.

To explore the behavior of this model, 
we have solved Eq.(\ref{EOM1d}) numerically with initial condition (\ref{IC1d}) 
for giant spins with $I = 25$, $I = 50$, and $I = 100$. 
The polarization transfer $\Avg{\Delta m}$, Eq.(\ref{DeltaM1d}), is plotted for each initial polarization $m_0$ in the upper panel of Fig.\ref{Simulation}b. 
In addition, we also show the inverse of the average dwell time $\bar\tau = -\int_0^\infty t\, \frac{d}{dt}\Amp{\psi}{\psi} \, dt$ for each case in the bottom panel of Fig.\ref{Simulation}b.


Two very different situations arise depending on the relationship between the spin-orbit coupling matrix element $u$ and the maximum hyperfine coupling matrix element $v_{\rm max}$. 
In case I, indicated by the dotted line $u = u_{\rm I} > v_{\rm max}$ in Fig.\ref{Simulation}a, spin-orbit coupling dominates the dynamics for any initial polarization $m_0$, and nuclear spin pumping is strongly suppressed (see Fig.\ref{Simulation}b). 
For $u = u_{\rm II} < v_{\rm max}$, however, the dynamics can be dominated either by hyperfine coupling or by spin-orbit coupling, depending on the value of the initial polarization. 
The system possesses critical points $m = \pm m_*$, with\cite{LargeI} 
\begin{equation} 
\label{m*} m_* \approx \pm I\sqrt{1 - \left(\frac{u}{v_{\rm max}}\right)^2}, 
\end{equation} 
where, locally, the strengths of hyperfine and spin-orbit coupling are nearly equal. 
For initial polarizations satisfying $|m_0| < m_*$, the dynamics are hyperfine-dominated and approximately one unit of angular momentum is transferred to the nuclear spin subsystem per electron. 
Outside the critical points, i.e. for $|m_0| > m_*$, spin-orbit coupling dominates and polarization transfer is strongly suppressed (see Fig.\ref{Simulation}b). 
As the length of the giant spin increases, the distinction between the behaviors in these two regimes becomes more sharply defined. 
In particular, $\Avg{\Delta m}$ becomes sharply quantized to 1 for $|m_0| < m_*$. 
Near the critical points, the dwell time of an electron in the system can become very long (see lower panel of Fig.\ref{Simulation}b). 

 
The behavior in all of these regimes 
can be understood in terms of a simplified model. 
If the quantum walk (\ref{EOM1d}) only explores a window of sites which is small compared with the scale 
of the giant spin, $2I + 1$, then we may approximate the $m$-dependent hopping amplitudes $v_m$ by a single amplitude $v = v_{m_0}$ which characterizes the strength of the transverse hyperfine field when the giant spin has $z$-projection $m_0$. 
In this same spirit, we also extend the lattice to infinity, $-\infty < m < \infty$. 
These approximations make the quantum walk 
translationally invariant, and allow us to find an exact analytical solution to the dynamics. 
 
The translationally-invariant model described above is identical to the one dimensional non-Hermitian quantum walk which was studied in Ref.[\onlinecite{1DPRL}]. 
In 
that work, we found that the expected displacement $\Avg{\Delta m}$ achieved 
before decay 
is quantized as an integer: 
\begin{equation} 
\Avg{\Delta m} = \left\{\begin{array}{lr}1\quad & v > u \\ 0\quad & u > v\end{array}\begin{array}{c}\\ \!\!\!\!.\end{array}\right. 
\end{equation}  
The quantized value of $\Avg{\Delta m}$ is determined by the winding of the phase between two components of the Bloch eigenstates of the bipartite one-dimensional system as the momentum $k$ is taken through the Brillouin zone. 
Equivalently, the value of $\Avg{\Delta m}$ can be determined directly from the winding number of the complex amplitude $A_k = u + ve^{ik}$. 
In the regime where this winding number is zero, corresponding to the situation where spin-orbit coupling $u$ is stronger than the hyperfine coupling $v$, the result $\Avg{\Delta m} = 0$ indicates that no angular momentum is pumped into the nuclear spin subsystem. 
As we will see below, this behavior is quite general and persists for more refined models which go beyond the giant spin approximation. 
 
The behavior of the giant spin model, displayed in Fig.\ref{Simulation}b, closely resembles the prediction of the translationally-invariant model. 
Some distortions are observed in the highly polarized regions, $|m| \approx I$, where $v_m$ varies strongly with $m$. 
The striking suppression of decay near the upper critical point can be traced to the divergence of the lifetime which accompanies the topological transition between winding and non-winding phases in the translationally invariant model (see Ref.\onlinecite{1DPRL}). 
The topological transition is manifested in the non-translationally invariant system through the presence of a topologically-protected ``dark'' edge state which is localized at the phase boundary between winding and non-winding phases.  
This state has zero overlap with the electron singlet state, and thus does not decay. 
Physically, the extended lifetime results from the fact that, near the critical point, the effective hyperfine and spin-orbit fields responsible for electron spin transitions can cancel each other. 
Although suppression of decay is seen near both critical points, the effect is much more dramatic near the upper critical point where the dark state is stable; near the lower critical point, an analogous exponentially-growing (delocalized) dark state can be found. 
 
In addition to the behavior of the lifetime, the small ``overshoots'' in $\Avg{\Delta m}$ can also be understood with intuition gained from the translationally-invariant model. 
Semiclassically, the site-dependent hopping amplitudes $v_m$ give the ``walker'' a position-dependent effective mass which goes through a minimum at the critical point. 
As a result, the walker experiences a force which attracts it to the critical point. 
For $m_0 \lesssim m_*$, where each electron already has a tendency to transfer one unit of angular momentum to the nuclear spin subsystem, this additional force results in a transfer of more than one unit of angular momentum per electron, $\Avg{\Delta m} > 1$.\cite{LCons} 
Similarly, above the critical point, the attraction leads to a negative polarization, $\Avg{\Delta m} < 0$. 
The shape of $\Avg{\Delta m}$ around the lower critical point can be understood analogously. 
 
\begin{figure} 
\includegraphics[width=3.25in]{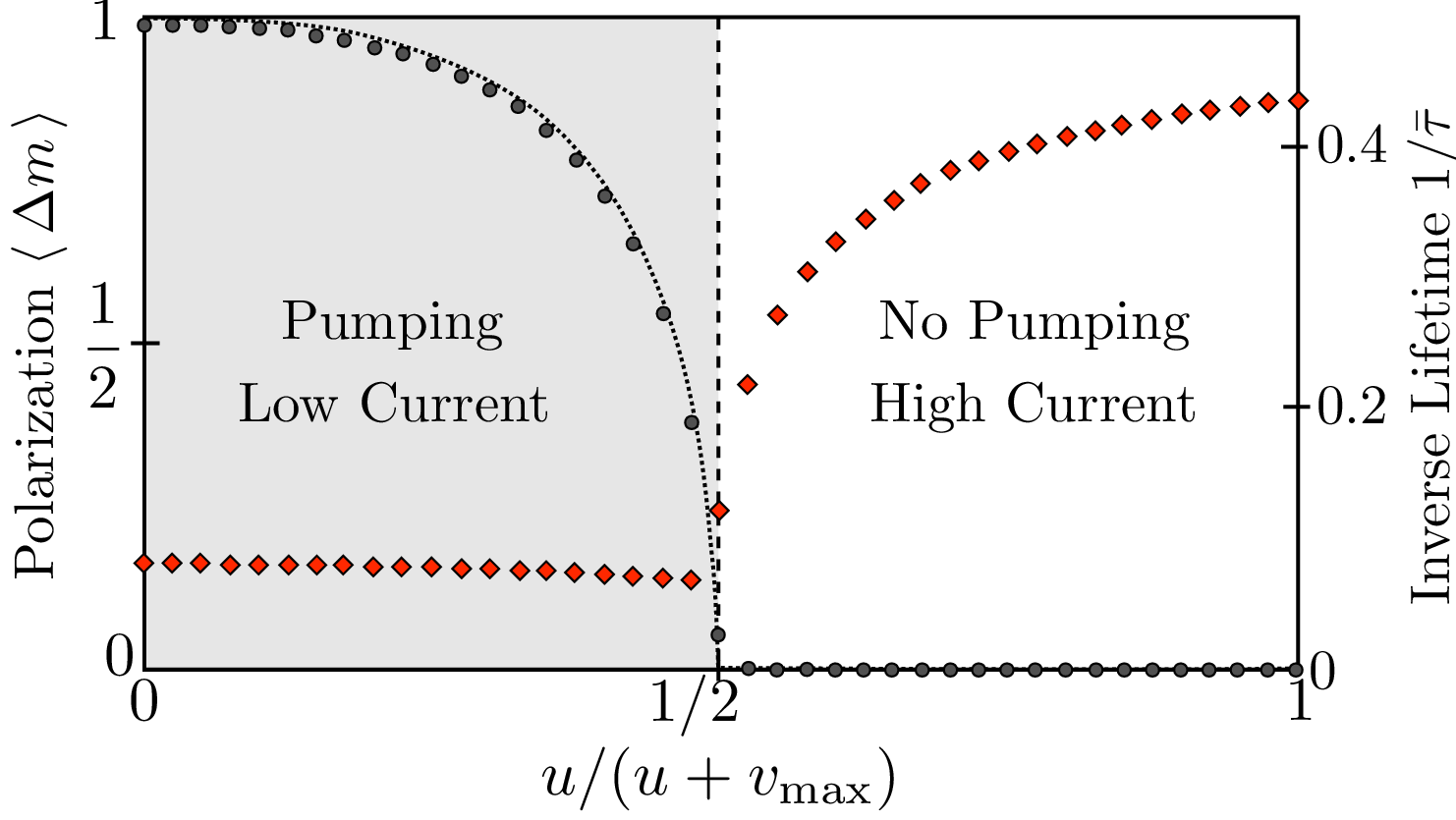}
 \caption[]{ Phase transition between pumping and non-pumping regimes for nuclear polarization described by the giant spin model.
Average polarization change $\overline{\Delta m}$ (circles) and lifetime $\bar\tau$ (diamonds) averaged over all initial states $-I \le m_0 \le I$ {\it vs.} the  ratio of the hyperfine and spin-orbital couplings.
The 
black line shows the average displacement $\overline{\Avg{\Delta m}}$, obtained by applying the translationally-invariant quantum walk model around each initial condition, Eq.(\ref{AvgMFormula}). 
} 
\label{SimulationAvg} 
\end{figure}

In experiments, the length $I$ of the giant spin and its $z$-projection $m_0$ in the initial state are in general random variables picked from the thermal distribution; the length $I$ is distributed according to $p(I) \propto (2I+1)^2\,e^{-(I/I_0)^2}$, with $I_0$ a constant, while $m_0$ is uniformly distributed for each choice of $I$, $p(m_0) = {\rm const}$. 
This formula for $p(I)$ is obtained for spin-1/2 nuclei using Eq.(1) of Ref.[\onlinecite{Mikhailov}] and Stirling's formula. 
In Fig.\ref{SimulationAvg} we show the numerically-obtained expected polarization transfer $\Avg{\Delta m}$ and inverse lifetime $1/\bar{\tau}$, 
averaged over all initial conditions $m_0$, for a giant spin with $I = 50$.  
The existence of dark states near the critical points 
is manifested in the suppressed current for 
$u < v_{\rm max}$. 
Compared with the translationally-invariant case, where $\Avg{\Delta m}$ displays a sharp step as a function of $u/(u + v)$, here the step is 
rounded into the phase where $v \lesssim v_{\rm max}$.  
When spin-orbit coupling dominates, however, i.e. for $u > v_{\rm max}$, the suppression of nuclear spin pumping is nearly complete. 
 
The shape of the rounded step can also be understood simply within the context of the translationally-invariant quantum walk model. 
For each initial condition $m_0$, 
we take $\Avg{\Delta m} = 1$ if $|m_0| < m_*$ (hyperfine dominates), or $\Avg{\Delta m} = 0$ if $|m_0| > m_*$ (spin-orbit dominates). 
After averaging over all initial polarizations $m_0$, we find 
\begin{eqnarray} 
  \label{AvgMFormula} \overline{\Avg{\Delta m}} &=& 0\times\left(1 - \frac{m_*}{I}\right) + 1\times\frac{m_*}{I} = \frac{m_*}{I}, 
\end{eqnarray} 
where $m_*$ is given by Eq.(\ref{m*}). 
Expression (\ref{AvgMFormula}) is plotted as a dotted line in Fig.\ref{Simulation}b. 
The good agreement with exact numerics for the giant spin model further indicates that the intuition gained from the translationally-invariant model provides a very useful tool for understanding the behavior in the more realistic situation.

\section{Beyond Giant Spin ($d > 1$) \label{sec:generalCase}} 
 
Faced with the surprising prediction of complete suppression of DNP in the spin-orbit dominated phase, it is natural to wonder to what extent this result relies on the assumption of uniform hyperfine coupling (i.e. on the validity of the giant spin model). 
The giant spin approximation tightly constrains the dynamics by truncating the dimension of the Hilbert space from $2^N$ down to $\mathcal{O}(\sqrt{N})$, where $N$ is the number of nuclear spins in the system. 
This appears to be a rather severe approximation. 
To address this concern, we now explore a more general class of models, derived in a similar spirit, which allow us to investigate the effects of nonuniform coupling. 

As illustrated in Fig.\ref{Cartoons}c, hyperfine coupling is generally strong near the center of the dots, where electron density is maximal, and weak near the edges, where electron density is small. 
To improve upon the uniform coupling model, consider dividing 
the nuclei into two groups, one of ``strongly-coupled'' spins, and the other of ``weakly-coupled'' spins. 
In this approximation, depicted in Fig.\ref{MultiDQW}b, the nuclear spins within each group 
form two separately conserved collective spins of lengths $I_1$ and $I_2$. 
Here the coupled dynamics of electron and nuclear spins can be viewed as a hopping model on a two dimensional lattice, where the two dimensions index the $z$-projections of the two collective spins, $m_1$ and $m_2$. 
This model captures both the transfer of polarization from the electron spins to the nuclear spins, and the RKKY-like electron-mediated transfer of polarization between the two groups of nuclear spins\cite{KLG02,Yao06,Deng06,Cywinski09}.

 
Continuing this reasoning, we can further refine the model by approximating the smooth electron density profile by $d$ shells of constant density $\rho_\alpha$, with $\alpha = 1 \ldots d$ (see red dotted line in Fig.\ref{Cartoons}c, and e.g. Ref.[\onlinecite{Gullans}]). 
In this case, the nuclear spins couple to form $d$ collective spins $I_1 \ldots I_d$. 
The corresponding polarization transfer dynamics can be viewed as a hopping problem on a $d$-dimensional lattice indexed by the polarizations $m_1, \ldots, m_d$ of the $d$ collective spins (see Fig.\ref{MultiDQW}c). 
This ``quantum walk'' is described by equations of motion analogous to Eq.(\ref{EOM1d}), with the position variable $m$ replaced by a vector ${\bf m} = (m_1, \ldots, m_d)$: 
\begin{eqnarray} 
  \label{EOMd}  
\begin{array}{rcrcccl} 
 i \, \dot{\psi}^T_{\vec{m}} &=&  \Delta\varepsilon\, \psi^T_{\vec{m}} &+&  u\, \psi^S_{\vec{m}} & +&  \sum_{\alpha}v_{\vec{m}}^{(\alpha)} \psi^S_{\vec{m}+\vec{e}_\alpha}\\ 
i \, \dot{\psi}^S_{\vec{m}} &=& -i\gamma/2\, \psi^S_{\vec{m}} &+& u\, \psi^T_{\vec{m}}  &+& \sum_{\alpha} v_{\vec{m}-\vec{e}_\alpha}^{(\alpha)} \psi^T_{\vec{m}-\vec{e}_\alpha}, 
\end{array} 
\end{eqnarray} 
where $v_{\vec{m}}^{(\alpha)} = A_\alpha\sqrt{I_\alpha(I_\alpha + 1) - m_\alpha(m_\alpha + 1)}$ is the hyperfine coupling to collective spin $I_\alpha$, with $A_\alpha = A\rho_\alpha$, and $\vec{e}_\alpha$ is the unit vector along the axis describing the polarization of $I_\alpha$. 
 
Our goal is now to calculate the polarization $\Avg{\Delta \vec{m}}$ transferred into the nuclear spin bath, 
\begin{eqnarray} 
  \label{DeltaM}  \Avg{\Delta \vec{m}} \equiv \sum_m (\vec{m} - \vec{m}_0)\, P_{\vec{m}}  
,\ 
P_{\vec{m}}=\int_0^\infty \gamma | \psi^S_{\vec{m}}(t)|^2 \, dt, 
\end{eqnarray} 
under the dynamics of Eq.(\ref{EOMd}) with initial condition 
\begin{eqnarray} 
  \label{IC} \psi^T_{\vec{m}} = \delta_{\vec{m},\vec{m}_0}, \quad  \psi^S_{\vec{m}} = 0. 
\end{eqnarray} 
In particular, we will be interested in determining which features of the results are independent of the particular grouping into collective spins, and which survive as the level of refinement, $d$, is increased. 

Based on the success of the translationally-invariant approximation to the quantum walk in the giant spin model (i.e. the case $d = 1$),  we begin by replacing the $\vec{m}$-dependent hopping amplitudes $v^{(\alpha)}_{\vec{m}}$ by constants $v^{(\alpha)} = v^{(\alpha)}_{\vec{m}_0}$. 
Note that the approximations associated with making hopping translationally invariant and with extending the lattice of states for each collective spin to infinity 
become more severe as the size of each spin decreases.  
Thus, although we will proceed for arbitrary dimension $d$, this number should be considered small compared to the total number of nuclear spins in the double dot. 
 
 
The next step is to pass to the momentum representation, 
$\psi^S_{\vec{m}} = \frac1{(2\pi)^d}\oint d^dk\, e^{i \vec{k}\cdot\vec{m}} \psi^S_{\vec{k}}$, where the integral is taken over the Brillouin zone $-\pi \le k_\alpha < \pi$, with $\alpha \in \{1, 2, \ldots, d\}$. 
These Fourier states correspond to coherent nuclear spin states 
with the transverse component of each collective spin $\alpha$ pointing along the azimuthal angle 
$k_\alpha$. 
Due to the translational invariance of the system, 
the equations of motion in the Fourier representation break up into $2 \times 2$ blocks, one for each momentum $\vec{k}$: 
\begin{eqnarray} 
  \label{EOMk} i \hbar \, \frac{d}{dt} 
  \left( 
    \begin{array}{c} 
      \psi^T_{\vec{k}} \\ 
      \psi^S_{\vec{k}} 
    \end{array} 
  \right) 
   = 
  \left( 
     \begin{array}{cc} 
       \varepsilon_T & A_{\vec{k}} \\ 
       A_{\vec{k}}^* & \tilde{\varepsilon}_S 
     \end{array} \right) 
  \left( 
    \begin{array}{c} 
      \psi^T_{\vec{k}} \\ 
      \psi^S_{\vec{k}} 
    \end{array} 
  \right), 
\end{eqnarray} 
with $A_{\vec{k}} = u + \sum_{\alpha=1}^{d} v^{(\alpha)}\, e^{i k_\alpha}$. 
The two-component wave functions for different values of $\vec{k}$ evolve independently, and the probability density  $p_{\vec{k}}(t) \equiv |\psi_{\vec{k}}^T(t)|^2 + |\psi_{\vec{k}}^S(t)|^2$ to find the system with momentum $\vec{k}$ at time $t$ decays as $\partial_t \, p_{\vec{k}} = -\gamma |\psi_{\vec{k}}^S(t)|^2$. 
The $\vec{k}$-dependence of $|A_{\vec{k}}|$ 
indicates that for some giant spin configurations ($k_\alpha \approx 0$) the effective hyperfine and spin-orbit fields add constructively, while for other configurations ($k_\alpha \approx \pi$) they interfere destructively. 

Writing $m_\alpha$ as a derivative with respect to $k_\alpha$ via $ m_\alpha\, \psi_{\vec{m}}^S = -\frac{i}{(2\pi)^d}\oint d^dk \, \frac{d}{dk_\alpha}\left(e^{i \vec{k}\cdot\vec{m}}\right) \psi^S_{\vec{k}}$ and integrating by parts to move the derivative onto $\psi^S_{\vec{k}}$, we bring the expression for the $\alpha$-th component of $\Avg{\Delta \vec{m}}$, Eq.(\ref{DeltaM}), to the form 
%
\begin{eqnarray} 
\!\!\!\!\!\!\!  \Avg{\Delta m_\alpha} &=& i\gamma \int_0^\infty dt \oint \frac{d^dk}{(2\pi)^d}\, {\psi^S_{\vec{k}}}^* \, \frac{\partial \psi^S_{\vec{k}}}{\partial k_\alpha},\\ 
  \label{DeltaM_k}   &=&\! \oint\! \frac{d^{d-1}k}{(2\pi)^{d-1}}\!\left\{ i\gamma\! \int_0^\infty\!\!\! dt \oint \frac{dk_\alpha}{2\pi}\, {\psi^S_{\vec{k}}}^* \, \frac{\partial \psi^S_{\vec{k}}}{\partial k_\alpha}\right\}\! . 
\end{eqnarray} 
The outer integral is taken over the $d-1$ momenta $k_{\beta \neq \alpha}$.  
The expression inside the braces is identical to Eq.(5) of Ref.[\onlinecite{1DPRL}] for the displacement in the one-dimensional model. 
As shown there, the value of this integral is quantized as either 0 or 1 depending on the winding of $\theta_{\vec{k}} \equiv {\rm arg}\{A_{\vec{k}}\}$ as $k_\alpha$ is taken around the Brillouin zone. 
 
For the one-dimensional case, quantization means that the expected change in polarization per electron through the system is either 1 if $\theta_k$ wraps the origin ($v > u$, hyperfine coupling exceeds spin-orbit coupling) or 0 if it does not ($u > v$, spin-orbit coupling exceeds hyperfine coupling). 
Roughly speaking, the winding of $\theta_{k}$ therefore distinguishes whether or not the hyperfine coupling is strong enough for the electron to flip the nuclear spin. 
 
\begin{figure}[t] 
\center{\includegraphics[width=3.25in]{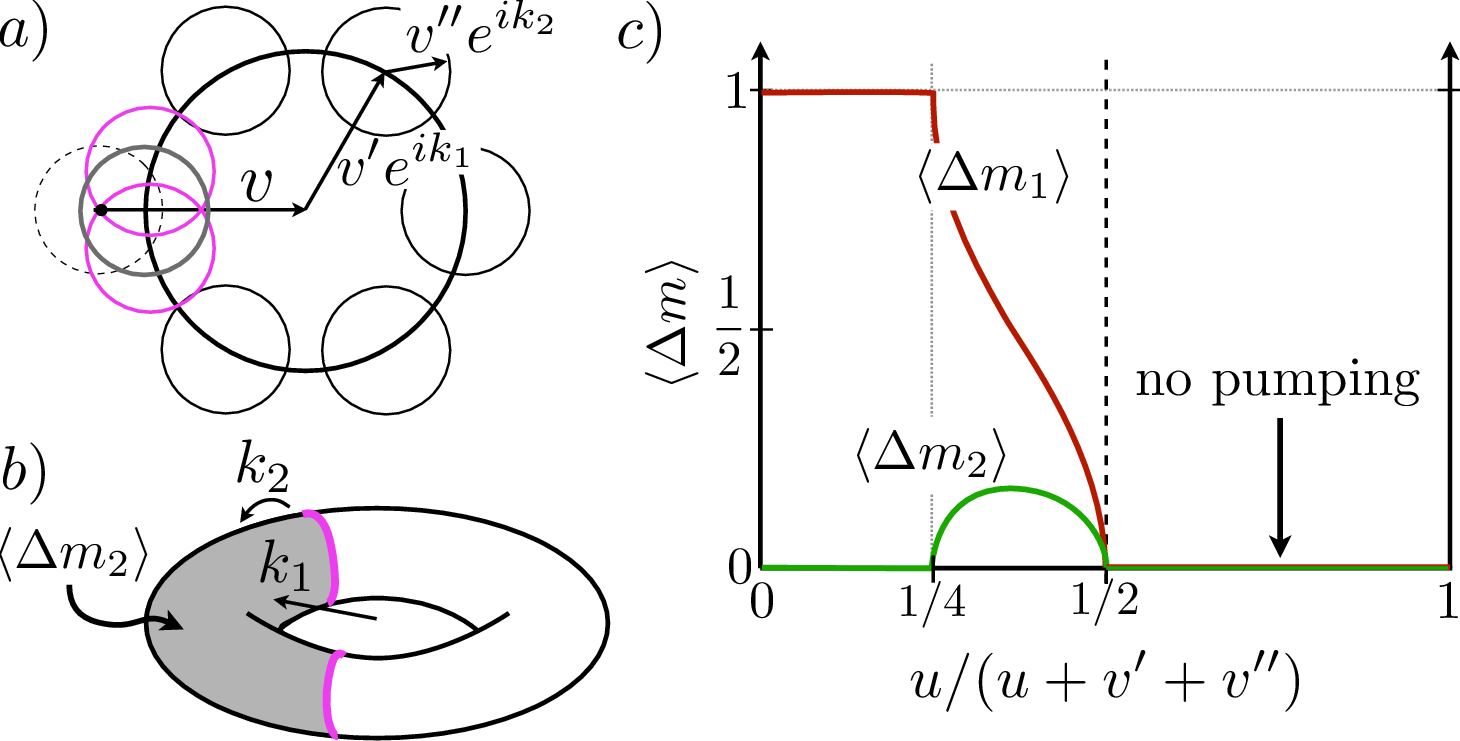}} 
 \caption[Graphical construction for evaluating mean displacements in d-dimensions]{Expected polarization from $d = 2$ quantum walk model. 
a) Graphical analysis of Eq.(\ref{DeltaM_k}). 
For each fixed $k_1$, $A_{\vec{k}} = u + v'e^{ik_1} + v''e^{ik_2}$ sweeps out a circle as $k_2$ is varied from $-\pi$ to $\pi$, encircling the origin for a range around $k_1 \approx -\pi$ bounded by the magenta circles. 
b) 
The mean displacement $\langle \Delta m_2\rangle$ is equal to the fraction of the Brillouin zone in which, for fixed $k_1$, $A_{\vec{k}}$ wraps the origin as $-\pi \le k_2 < \pi$. 
A similar construction can be used to obtain $\langle \Delta m_1\rangle$ (not shown).
c) Solution to Eq.(\ref{DeltaM_k}) for $\Avg{\Delta m_1}$ and $\Avg{\Delta m_2}$ versus the ratio of spin-orbit and hyperfine coupling strengths. 
The ratio between the two giant spin hyperfine couplings 
is fixed to the value $v'/v'' = 2$. 
When spin-orbit coupling $u$ is weak, the more strongly coupled giant spin absorbs all spin-flips ($\Avg{\Delta m_1} = 1, \Avg{\Delta m_2} = 0$). 
When spin-orbit coupling exceeds the maximum hyperfine coupling, $u > v' + v''$, nuclear spin pumping is completely suppressed. 
} 
\label{TPTGraphConst} 
\end{figure} 
 
To understand the meaning of Eq.(\ref{DeltaM_k}) in the multi-dimensional case, it is helpful to view the integral in the following way: for each fixed configuration of $d-1$ collective spins described by the $d-1$ angles $\{k_{\beta \neq \alpha}\}$, the expression inside the braces is either 0 or 1 depending on, for the given field of the other spins and strength of spin-orbit coupling, whether the electron's hyperfine coupling $v^{(\alpha)}$ to the remaining spin is strong enough to induce a spin flip (i.e. whether or not $\theta_{\vec{k}}$ winds the origin as $k_\alpha$ is varied from $-\pi$ to $\pi$, see Fig.\ref{TPTGraphConst}a). 
The integral over the remaining $d-1$ variables simply counts the ``phase space'' over which this condition is satisfied.  
This result is represented graphically for the case $d=2$ in Fig.\ref{TPTGraphConst}b. 
 
\subsection{Special Case: $d = 2$} 
Recently, the relative dynamics of nuclear spins in the two dots of spin-blockaded double quantum dots has attracted considerable experimental\cite{Reilly, FolettiControl} and theoretical\cite{RamonZamboni, Ribeiro, Fluctuations, StopaForce, Gullans} attention. 
From a practical point of view, understanding the behavior of the difference of polarization between the two dots is important because a) it is responsible for dephasing of singlet-triplet qubits, and b) because the polarization difference, if carefully controlled, can be used as a resource to coherently control electron spin dynamics\cite{FolettiControl}. 
The case of our model with $d = 2$, where the electron spins couple to two independent collective spins, is thus particularly interesting if the two spins are viewed as representing the nuclear spin states in the left and right dots. 
We analyze this case in detail in this subsection. 
 
The expected displacements $\Avg{\Delta m_1}$ and $\Avg{\Delta m_2}$ in the two-dimensional quantum walk, see Eq.(\ref{DeltaM_k}), represent the expected amounts of polarization transferred to each of the two groups of nuclear spins during the decay of the electron spin. 
The values of $\Avg{\Delta m_1}$ and $\Avg{\Delta m_2}$ depend on the strength of spin orbit coupling, $u$, and the strengths of the transverse hyperfine fields produced by the two giant spins, $v'$ and $v''$, respectively. 
Using simple geometric arguments based on the construction shown in Fig.\ref{TPTGraphConst}a, we find (assuming $v' > v''$) 
\begin{equation} 
\label{DeltaM2D}\Avg{\Delta m_1} =  
\left\{\begin{array}{l} 
    1\\ 
    1 - \frac{\theta_1}{\pi}\\
    0\\ 
\end{array}\right., \  
\Avg{\Delta m_2} =  
\left\{\begin{array}{lcr} 
    0 &\ \ & v' > u + v''\\ 
    \frac{\theta_2}{\pi}&\ \ & |u - v'| \le v''\\
    0&\ \ &v' < u - v'',\end{array}\right. 
\end{equation} 
with 
\begin{equation} 
\cos\theta_1 = \frac{v'^2 - u^2 - v''^2}{2uv''}, \quad \cos\theta_2 = \frac{v'^2 + u^2 - v''^2}{2uv'}. 
\end{equation} 
 
 
Expression (\ref{DeltaM2D}) is plotted in Fig.\ref{TPTGraphConst}c as a function of spin-orbit coupling strength $u$ for the fixed ratio $v'/v'' = 2$. 
For strong spin-orbit coupling $u > v' + v''$, neither collective spin is pumped at all: we find complete suppression of DNP in the spin-orbit dominated phase just as in the single giant spin model. 
Interestingly, for weak spin-orbit coupling we find a quantization $\Avg{\Delta m_1} = 1, \Avg{\Delta m_2} = 0$, which indicates that on average the collective spin with stronger hyperfine coupling to the electron absorbs the full angular momentum, while the more weakly coupled spin is not pumped at all. 
For an asymmetric system composed of one large dot and one small dot, electron density and thus average hyperfine coupling is larger on the smaller dot. 
Thus in the limit of weak spin-orbit coupling, $u \ll v',v''$, nuclear pumping may be highly asymmetric, with DNP initially produced primarily on the {\it smaller} of the two dots. 
 
 
\subsection{General Results} 
In contrast to the one-dimensional translationally-invariant model, Fig.\ref{TPTGraphConst}c shows that in higher dimensions $\Avg{\Delta m_\alpha}$ is not strictly quantized as an integer. 
The breakdown of quantization results from the appearance of ``mixed'' phases where the winding number along one dimension of the Brillouin zone is 1 for some values of the remaining momenta, and 0 for the others. 
However, for the quantum walk (\ref{EOMd}) in any dimension $d$ there is {\it always} a ``non-winding'' phase with $u > \sum_\alpha v^{(\alpha)}$, where all winding numbers are 0 for all values of $\vec{k}$ in the Brillouin zone. 
In this phase, $\Avg{\Delta m_{\alpha}} = 0$ for all $\alpha$. 
In fact, either through graphical methods or with a few lines of algebra, one can see that, as $u$ is increased for fixed $\{v^{(\alpha)}\}$, {\it all} of the $\Avg{\Delta m_\alpha}$ vanish simultaneously at the point $u = \sum_\alpha v^{(\alpha)}$. 
Thus very generally, strong spin-orbit coupling can lead to a dramatic suppression of nuclear spin pumping. 
 
Furthermore, 
the expected polarizations $\Avg{\Delta m_\alpha}$ are determined purely by geometrical constraints imposed by the set of hyperfine matrix elements $v^{(\alpha)}$. 
Importantly, just as in the 1d case studied in Ref.[\onlinecite{1DPRL}], both the detuning $\epsilon_T^0 - \epsilon_S$ and the decay rate $\gamma$ completely drop out of the solution. 
As shown there, 
the result holds even if these quantities are made time-dependent. 
This implies that the suppression of pumping results from nuclear spin coherence, and is robust against noise and dephasing of the electron singlet and triplet states. 

\section{Conclusions} 
In this work we have analyzed
the coupled dynamics of electron and nuclear spins in spin blockaded quantum dots by mapping it
onto a non-Hermitian quantum walk. 
This quantum walk problem can be solved exactly by making the approximation that the matrix elements of the hyperfine coupling are roughly independent of the nuclear polarization $m$ for $|m - m_0| \ll I$, where $I$ is the size of a large collective spin formed from a group of nuclei with similar values of the hyperfine coupling to the electron. 
From this solution we find that nuclear spin pumping is strongly suppressed whenever the strength of spin-orbit coupling exceeds the magnitude of the hyperfine coupling. 
Numerical simulations show that this behavior extends directly to the more realistic situation where the polarization-dependence of the hyperfine matrix elements is included. 
 
The transition in the nuclear spin pumping efficiency is accompanied by an abrupt change in average current through the double dot. 
Due to the presence of ``dark states'' when the hyperfine and spin-orbit processes exhibit complete destructive interference,  
the expected lifetime of the spin-blockaded triplet state is significantly longer in the regime where hyperfine and spin-orbit couplings compete than in a regime where only one of the two mechanisms is present. 
This interference is mediated by coherence in the nuclear spin bath, and is therefore robust against dephasing of the electronic state.

We thank B. I. Halperin, J. Krich, and I. Neder for helpful discussions, and acknowledge  
support from W. M. Keck 
Foundation Center for Extreme Quantum Information 
Theory, from  
NSF Grants DMR-090647 and PHY-0646094 (M.R.) and from 
NSF Grant No. PHY05-51164 (L. L.). 
 
 

\begin{references} 
 
\bibitem{OnoSB} 
K. Ono, D. G. Austing, Y. Tokura, S. Tarucha, Science {\bf 297}, 1313 (2002). 
 
\bibitem{Koppens} 
F. H. L. Koppens, J. A. Folk, J. M. Elzerman, R. Hanson, L. H. Willems van Beveren, I. T. Vink, H. P. Tranitz, W. Wegscheider, L. P. Kouwenhoven, L. M. K. Vandersypen, Science {\bf 309}, 1346-1350 (2005).  
 
\bibitem{Liu08} 
H. W. Liu, T. Fujisawa, Y. Ono, H. Inokawa, A. Fujiwara, K. Takashina, Y. Hirayama, Phys. Rev. B {\bf 77}, 073310 (2008). 
 
\bibitem{Shaji08} 
N. Shaji, C. B. Simmons, M. Thalakulam, L. J. Klein, H. Qin, H. Luo, D. E. Savage, M. G. Lagally, A. J. Rimberg, R. Joynt, M. Friesen, R. H. Blick, S. N. Coppersmith, M. A. Eriksson, Nature Physics {\bf 4}, 540 (2008). 
 
\bibitem{Pfund} 
A. Pfund, I. Shorubalko, K. Ensslin, R. Leturcq, Phys. Rev. Lett. {\bf 99}, 036801 (2007).  
 
\bibitem{Frolov} 
S. Nadj-Perge, S. M. Frolov, J. W. W. van Tilberg, J. Danon, Yu. V. Nazarov, R. Algra, E. P. A. M. Bakkers, L. P. Kouwenhoven, arXiv:1002.2120. 
 
 
\bibitem{Churchill} 
H. O. H. Churchill, A. J. Bestwich, J. W. Harlow, F. Kuemmeuth, D. Marcos, C. H. Stwertka, S. K. Watson, and C. M. Marcus, Nature Physics {\bf 5}, 321 (2009). 
 
 
 
\bibitem{OnoTarucha} 
K. Ono and S. Tarucha, Phys. Rev. Lett. {\bf 92}, 256803 (2004).  
 
\bibitem{RudnerDNP} 
M. S. Rudner and L. S. Levitov, Phys. Rev. Lett. {\bf 99}, 036602 (2007). 
 
\bibitem{RudnerCooling} 
M. S. Rudner and L. S. Levitov, arXiv:0705.2177. 
 
\bibitem{Larmor} 
M. S. Rudner, I. Neder, L. S. Levitov, and B. I. Halperin, arXiv:0909.0060. 
 
\bibitem{Taylor03} 
J. M. Taylor, A. Imamoglu, and M. D. Lukin, 
Phys. Rev. Lett. {\bf 91}, 246802 (2003). 
 
\bibitem{1DPRL} 
M. S. Rudner and L. S. Levitov, Phys. Rev. Lett. {\bf 102}, 065703 (2009). 
 
\bibitem{LargeI} 
Here we use the fact that $I(I+1) \approx I^2$ for $I \gg 1$. 

\bibitem{LCons}
This result does not violate conservation of angular momentum, as the spin-orbit interaction allows the spins to exchange angular momentum with orbital degrees of freedom and eventually to the lattice.

\bibitem{Mikhailov} 
V. V. Mikhailov, J. Phys. A: Math. Gen. {\bf 10}, 147 (1977). 
 
\bibitem{Shenvi05} 
N. Shenvi, R. de Sousa, and K. B. Whaley, Phys. Rev. B {\bf 71}, 224411 (2005).   
 
\bibitem{KLG02} 
A. V. Khaetskii, D. Loss, and L. Glazman, Phys. Rev. Lett. {\bf 88}, 186802 (2002). 
 
\bibitem{Yao06} 
W. Yao, R.-B. Liu, and L. J. Sham, Phys. Rev. B {\bf 74}, 195301 (2006). 
 
\bibitem{Deng06} 
C. Deng and X. Hu, Phys. Rev. B {\bf 73}, 241303 (2006). 
 
\bibitem{Cywinski09} 
L. Cywinski, W. M. Witzel, S. Das Sarma, Phys. Rev. Lett. {\bf 102}, 057601 (2009). 
 
 
 
 
 
 
 
 
 
 
 
 
 
 
 
 
 
 
 
\bibitem{Gullans} 
M. Gullans, J. J. Krich, J. M. Taylor, H. Bluhm, B. I. Halperin, C. M. Marcus, M. Stopa, A. Yacoby, and M. D. Lukin,  arXiv:1003.4508. 
 
\bibitem{Reilly} 
D. J. Reilly, J. M. Taylor, J. R. Petta, C. M. Marcus, M. P. Hanson, and A. C. Gossard, Science {\bf 321}, 817 (2008). 
 
\bibitem{FolettiControl} 
S. Foletti, H. Bluhm, D. Mahalu, V. Umansky, and A. Yacoby, Nature Physics {\bf 5}, 903 (2009). 
 
\bibitem{RamonZamboni} 
G. Ramon and X. Hu, Phys. Rev. B. {\bf 75}, 161301(R) (2007). 
 
\bibitem{Ribeiro} 
H. Ribeiro and G. Burkard, Phys. Rev. Lett. {\bf 102}, 216802 (2009). 
 
\bibitem{Fluctuations} 
M. S. Rudner, F. H. L. Koppens, J. A. Folk, L. M. K. Vandersypen, and L. S. Levitov, arXiv:1001.1735. 
 
\bibitem{StopaForce} 
M. Stopa, J. J. Krich, and A. Yacoby, Phys. Rev. B. {\bf 81}, 041304(R) (2010). 
 
 
 
 
 
 
 
 
 
 
 
 
 
 
 
 
 
 
 
 
 
 
 
 
 
 
 
 
 
 
 
\end{references}
\end{document}